\documentclass[%
 reprint,
superscriptaddress,
 amsmath,amssymb,
 aps,
]{revtex4-1}

\usepackage{graphicx}% Include figure files
\usepackage{dcolumn}% Align table columns on decimal point
\usepackage{bm}% bold math
\usepackage{tensor}
\usepackage{svg}
\usepackage{subcaption}
\usepackage{physics}
\usepackage[%  
    colorlinks=true,
    pdfborder={0 0 0},
    linkcolor=blue
]{hyperref}
\usepackage{url}
\begin{document}

\preprint{APS/123-QED}

\title{Robust millisecond coherence times of erbium electron spins}% Force line breaks with \\

% \affiliation{Department of Physics, University of Chicago, Chicago IL}

% \affiliation{Corning Research and Development Corporation, New York }

% \affiliation{Pritzker School of Molecular Engineering, University of Chicago, Chicago IL}

\author{Shobhit Gupta}
%  \homepage{http://www.Second.institution.edu/~Charlie.Author}
\affiliation{Department of Physics, University of Chicago, Chicago IL 60637 USA}

%  Second institution and/or address\\
%  This line break forced% with \\
% }%
% \affiliation{
%  Third institution, the second for Charlie Author
% }%
\author{Xuntao Wu}
\affiliation{Pritzker School of Molecular Engineering, University of Chicago, Chicago IL 60637 USA}
%  Authors' institution and/or address\\
%  This line break forced with \textbackslash\textbackslash

\author{Haitao Zhang }
\affiliation{Corning Research and Development Corporation, Sullivan Park, Painted Post, NY 14870, USA  }

\author{Jun Yang}
\affiliation{Corning Research and Development Corporation, Sullivan Park, Painted Post, NY 14870, USA }
%  Authors' institution and/or address\\
%  This line break forced with \textbackslash\textbackslash
%\author{Jun Yang}
%\affiliation{Corning Research and Development Corporation, New York}
% }%
\author{Tian Zhong}
\affiliation{Pritzker School of Molecular Engineering, University of Chicago, Chicago IL 60637 USA}

\date{\today}% It is always \today, today,

\begin{abstract}
Erbium-doped solids are prime candidates for optical quantum communication networks due to erbium's telecom C-band emission. A long-lived electron spin of erbium with millisecond coherence time is highly desirable for establishing entanglement between adjacent quantum repeater nodes while long-term storage of the entanglement could rely on transferring to erbium's second-long coherence nuclear spins. Here we report GHz-range electron spin transitions of $^{167}\mathrm{Er}^{3+}$ in yttrium oxide ($\mathrm{Y_2O_3}$) matrix with coherence times that are consistently longer than a millisecond. Instead of addressing field-specific Zero First-Order Zeeman transitions, we probe weakly mixed electron spins with the field orientation along the lower g-factors. Using pulsed electron spin resonance spectroscopy, we find paramagnetic impurities are the dominant source of decoherence, and by polarizing them we achieve a Hahn echo spin $\mathrm{T_2}$ up to 1.46 ms, and a coherence time up to 7.1 ms after dynamical decoupling. These coherence lifetimes are among the longest found in crystalline hosts  especially those with nuclear spins. We further enhance the coherence time beyond conventional dynamical decoupling, using customized sequences to simultaneously mitigate spectral diffusion and Er-Er dipolar interactions. Despite nuclear and impurity spins in the host, this work shows that long-lived erbium spins comparable to non-nuclear spin hosts can be achieved. Our study not only establishes $^{167}\mathrm{Er}^{3+}$: $\mathrm{Y_2 O_3}$ as a significantly promising quantum memory platform but also provides a general guideline for engineering long-lived erbium spins in a variety of host materials for quantum technologies.
\end{abstract}

\maketitle
%\tableofcontents

\raggedbottom

\section{\label{sec:level1}Introduction}

Rare-earth-ion dopants in solids exhibit exceptionally long spin coherence times and narrow optical linewidths  \cite{THIEL2011353} \cite{Rancic2018} \cite{Zhong2015}, making them a prime candidate for quantum memories and quantum transduction \cite{pnas.1419326112} in an optical quantum network \cite{Kimble2008}, and with a good prospect for quantum sensing \cite{RevModPhys.89.035002}. Kramers rare-earth-ions with unpaired 4f electrons possess effective electron spin 1/2, and $\mathrm{Er}^{3+}$ in particular demonstrates telecom-wavelength optically addressable spins that are ideal for realizing quantum spin-photon interfaces in a fiber-based quantum communication network \cite{Raha2020, Asadi2020}. For such a network, the Er spin coherence time needs to be at least a millisecond in order to efficiently generate and store entanglement over the photon transit time between adjacent network nodes, with a typical repeater node interval of 100-200 km \cite{repeaterQM}. Once generated, the entanglement between remote Er spins can be transferred to $^{167}\mathrm{Er^{3+}}$ nuclear spin registers for long-term storage while the electron spins are freed up for subsequent entanglement generation. 

Due to large magnetic dipole moment, however, $\mathrm{Er}^{3+}$ electron spins experience strong decoherence from magnetic noise created by either $\mathrm{Er}^{3+}$ or other paramagnetic spin flip-flops, limiting their coherence to microseconds range in most prior reports \cite{multimodeeryso} \cite{eropticallyexcitedstate} \cite{merkel2020dynamical}  \cite{2107.04498}. Milliseconds $\mathrm{Er}^{3+}$ spin coherence time has only been demonstrated recently in bulk single crystals with low nuclear spin density (e.g. CaWO$_4$, 20 ppm, $\mathrm{T_{2} =  1.3 \,\, ms}$ \cite{https://doi.org/10.48550/arxiv.2203.15012}; undoped, $\mathrm{T_{2} =  23 \,\, ms}$ \cite{doi:10.1126/sciadv.abj9786}) at milliKelvin temperatures. Another approach to prolonging the $\mathrm{Er}^{3+}$ spin coherence considers $^{167}\mathrm{Er}^{3+}$ isotope, which possesses 7/2 nuclear spin with long-lived hyperfine transitions \cite{Rancic2018}. This gives rise to hybridized electron-nuclear spins \cite{Ortu2018} \cite{spinechogroundandopter} with Zero First-Order Zeeman transitions (ZEFOZ) showing reduced sensitivity to noise. Nevertheless ZEFOZ transitions often occur at zero or low fields, where the magnetic noise spectral density peaks due to unpolarized spin baths, and addressing them requires stringent field alignment, partially limiting the robustness and effectiveness of this method.

In this work, we report millisecond coherence times of electron spin transitions of $^{167}\mathrm{Er}^{3+}$ in yttrium oxide host. Instead of addressing ZEFOZ transitions, we probe weakly mixed electron spin transitions with low g-factors. By freezing the impurity spins and employing customized dynamic decoupling we demonstrate consistent Er spin coherence time over a millisecond threshold. We investigate 20 parts per million (ppm) doped $^{167}\mathrm{Er^{3+}}$ ensembles in $\mathrm{Y_2O_3}$ polycrystalline hosts using pulsed electron spin resonance (ESR) spectroscopy. Rather than host nuclear spins, we identify paramagnetic impurity spins as the primary source of decoherence. By freezing them at milliKelvin temperatures, we obtain Hahn echo \cite{hahnecho} coherence time of 1.46 ms, and $\mathrm{T_2^{\rm XY8}}$ of 7.1 ms after dynamical decoupling (DD) of spectral diffusion. By further designing customized DD sequences to simultaneously mitigate spectral diffusion and instantaneous diffusion we achieved a longer coherence than what is obtainable with conventional XY8 sequence. The strategies demonstrated in this work are applicable to a variety of host materials for erbium and other Kramers rare-earth spins, including those containing nuclear spin bath and paramagnetic impurities. Thus the study provides a general recipe for engineering long-coherence rare-earth spins as appealing resources for quantum technologies.

\begin{figure*}[th!]
    \centering
         \includegraphics[width=1\textwidth]{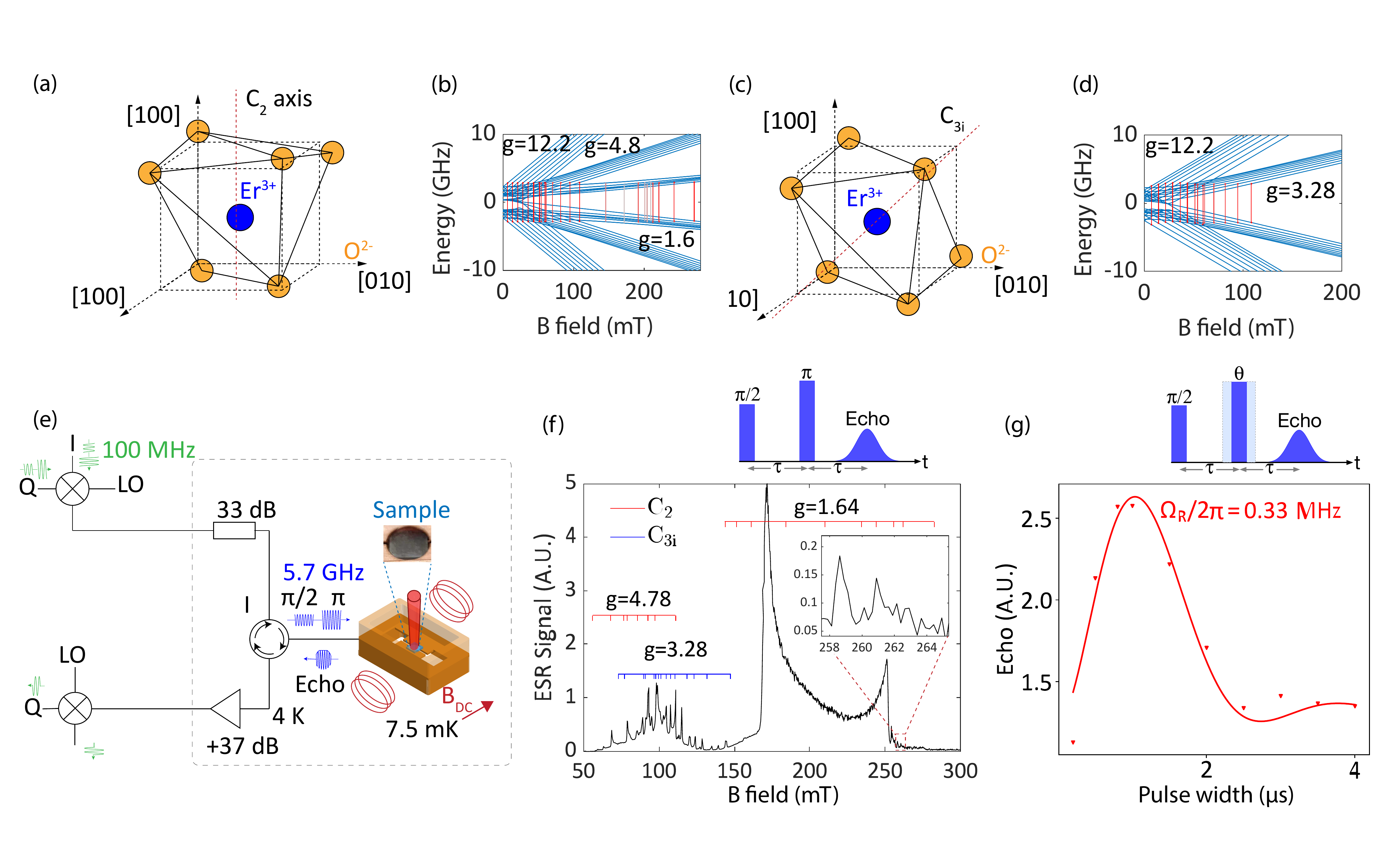}% first figure itself
    \caption{(a), (c) Er\textsuperscript{3+}:Y\textsubscript{2}O\textsubscript{3} crystal structure showing the unit cell and Er\textsuperscript{3+} substituting in C\textsubscript{2} and C\textsubscript{3i} sites, respectively. (b), (d) Zeeman and hyperfine energy levels for $^{167}\mathrm{Er}^{3+}$ in C\textsubscript{2} and C\textsubscript{3i} sites, respectively generated with Easyspin \cite{STOLL200642}. The red lines shows the allowed ESR transitions at 5.67 GHz resonance frequency. (e) MilliKelvin pulsed ESR setup. (f) Echo detected field sweep of $^{167}\mathrm{Er}^{3+}$ : Y\textsubscript{2}O\textsubscript{3} showing predicted transitions as dashes grouped by their g factors. (g) Rabi nutation of echo signal obtained by varying refocusing pulse length.}
\end{figure*}

\section{Experimental Methods}
We use yttrium oxide ($\mathrm{Y_2O_3}$) as a host matrix for $^{167}\mathrm{Er}^{3+}$ dopants, which can be prepared in diverse forms such as single crystals, polycrystalline ceramics, nanoparticles and thin-films \cite{nanoREreview}, with ultra-narrow, sub-kilohertz optical linewidths already demonstrated in $\mathrm{Er}^{3+}$ doped $\mathrm{Y_2O_3}$ transparent ceramics \cite{Zhang2017}\cite{yizhongandriku}. The sample under study is a 20 ppm $\mathrm{^{167}Er^{3+}}$ doped $\mathrm{Y_2O_3}$ ceramic with 95 $\mathrm{\%}$ isotopic purity prepared by sintering $\mathrm{Er^{3+}}$ doped $\mathrm{Y_2O_3}$ nanoparticles with grain dimensions of 0.7 - 2.1 $\mathrm{\mu m}$ and an average area of 0.90 $\mathrm{\mu m^2}$. Yttrium oxide ($\mathrm{Y_2O_3}$) has a cubic crystal structure ($T_h^7$ space group) with 16 formula units per unit cell \cite{harristhesis} \cite{reinemerthesis}. These 32 $\mathrm{Y^{3+}}$ sites can be grouped into two classes with 24 sites of $\mathrm{C_2}$ point group symmetry (Fig.~1(a)), and 8 sites for $\mathrm{C_{3i}}$ point group symmetry (Fig.~1(c)), where each of the 32 yttrium ($\mathrm{Y^{3+}}$) ions in a unit cell has been shown to be substituted with an $\mathrm{Er^{3+}}$ ion with an equal probability \cite{Sheller1} \cite{Schafer}.

$\mathrm{Er^{3+}}$ ion has a $\mathrm{[Xe]4f^{11}}$ configuration with $^{2S+1}\mathrm{L_{J}}=$  $^{4}\mathrm{I_{15/2}}$ ground state. In a crystal field of low symmetry the $\mathrm{J=15/2}$  ground state splits into 8 Kramers doublets with the lowest doublet being occupied at cryogenic temperatures, leading to an effective $\mathrm{S= 1/2}$. Erbium has 5 stable isotopes with zero nuclear spin $\mathrm{I=0}$, $^{162,164,166,168,170}\mathrm{Er}$, and a $^{167}\mathrm{Er}$ isotope with $\mathrm{I =7/2}$ nuclear spin with $\mathrm{22.95 \%}$ abundance. The $^{167}\mathrm{Er}$ isotope thus has 16 ground state hyperfine energy levels. 
 
The spin Hamiltonian of the lowest Kramers doublet of $\mathrm{J=15/2}$ manifold of $^{167}\mathrm{Er^{3+}}$ in $\mathrm{Y_2O_3}$ therefore can be modelled as
 
 \begin{equation}
     \mathcal{H} = \mu_e \boldsymbol{\mathrm{B\cdot g \cdot S}} + \boldsymbol{\mathrm{I\cdot A \cdot S}} + \boldsymbol{\mathrm{I\cdot Q \cdot I}} -\mu_n g_n \boldsymbol{\mathrm{B\cdot I}}
 \end{equation}

\noindent where $\mu_e$ is the Bohr magneton, $\boldsymbol{\mathrm{B}}$ is the external magnetic field, $\boldsymbol{\mathrm{g}} $ is the g-factor tensor, $\boldsymbol{\mathrm{A}} $ is the hyperfine tensor, $\mathrm{g_n}$ is the nuclear g factor, and  $\boldsymbol{\mathrm{Q}} $ is the nuclear quadrupole tensor.

%  In order to fit the spin Hamiltonian for the two sites, we performed continuous wave (cw-ESR) at X-band frequency on two $\mathrm{Er^{3+}}$ doped $\mathrm{Y_2O_3}$ samples- a 20 parts per million (ppm)  ceramic with natural isotopic abundance and a 20 ppm  ceramic with 95 $\mathrm{\%}$ purity of $^{167}\mathrm{Er^{3+}}$ (see Supplementary Material for fitting details).
 
%  We obtain for the $\mathrm{C_2}$ site the principle values of the $g$ tensor as $g_x=1.64$, $g_y=4.89$ and $g_z=12.21$, $A$ tensor as $A_x=212$, $A_y=548$ and $A_z=1288$ MHz. Similarly for the $\mathrm{C_{3i}}$ site  principle values of the $g$ tensor is $g_\parallel=3.32$, $g_\perp=12.2$, $A$ tensor is $A_\parallel=1267.78$, $A_\parallel=345$ MHz. A preliminary estimate of Q tensor was also obtained for both sites to explain forbidden transitions in cw-ESR (see Supplementary Material). A large hyperfine splitting A for both $\mathrm{C_2}$ and $\mathrm{C_{3i}}$ sites leads to an estimated zero field splitting of $\approx 4.6 \,\, \mathrm{GHz}$ corresponding to hybridized electron-nuclear states  \cite{Ortu2018} \cite{afzelius2}.
 
 We use the principle values of g, A and Q tensors for $\mathrm{C_2}$ and $\mathrm{C_{3i}}$ sites obtained from a separate study \cite{y2o3esr2022} to calculate the spin energy levels with magnetic field as shown in Fig.~1(b) and Fig.~1(d). Due to the random orientation of crystal grains in a polycrystalline sample, while ESR transitions from $\mathrm{g=12.2, g=4.78}$ and $\mathrm{g=3.28}$ overlap, we can individually address transitions with $\mathrm{g=1.64}$ in the region of 160-275 mT at 5.67 GHz ESR frequency  (Fig.~1(b) and Fig.~1(d)). The transitions with $\mathrm{g=1.64}$ are of interest in this study due to their lowest sensitivity to noise, and strong coupling to microwaves given the high transverse g factor up to 12.2 \cite{anisotropicg}. 
 
 We perform pulsed ESR on a sample with dimensions of $\mathrm{ 4.6 \, mm \, \times \, 4.6 \, mm \, \times 1 \, mm}$ , which is mounted in between the two halves of a 3D copper loop-gap resonator \cite{Hyde1989}. The resonator was made with oxygen-free copper and was assembled with two halves with a gap exactly matching the sample size for good thermal conduction (Fig.~1(e)). The resonator was mounted on the mixing chamber stage of a dilution refrigerator (Bluefors LD 250) and had a 5.67 GHz resonance frequency with a loaded Q of 3500 at 7.5 mK base temperature. A 3-axis vector magnet (AMI Model 430) was used to apply a DC bias field. Laser illumination was applied to the sample using an objective lens with weak focus on the sample with $\mathrm{\approx}$ 2 mm beamspot size on sample. 
 
%  The setup (Fig. \ref{setup}) generates phase controlled pulses at 5.67 GHz using IQ (in-phase and quadrature) upconversion of 100 MHz intermediate frequency (IF) pulses. The echo signal is preamplified using low noise amplifier (HEMT) on 4 Kelvin stage of the cryostat and downconverted to 100 MHz IF  signal detection using heterodyne technique.
 
3D resonators offer advantages over 2D superconducting microresonators for pulsed ESR, including higher field homogeneity, the ability to apply out-of-plane magnetic fields, easier integration with optics, as well as compatibility with diverse sample geometry \cite{SIMENAS2021106876}. However, reduced sensitivity of 3D resonators typically requires the use of high power pulses ($\mathrm{\approx Watts}$) leading to heating and elevated temperature ($\mathrm{\approx  \, 100 \, mK}$ \cite{milliKelvinESR}) of the ESR spectrometer. We use heterodyne detection technique along with a low noise preamplifier (HEMT) (Fig. 1e) to enhance the spectrometer sensitivity, allowing for low power ($\mathrm{ \approx \, 100 \, \mu W}$) pulsed ESR measurements. This results in a lowest possible temperature of the sample stage of 7.5 mK during the experiment, allowing us to maximally polarize spin population. Rabi frequency of 0.33 MHz is obtained in a two pulse echo sequence with varying refocusing (second) pulse length (Fig.~1(g)), which allows us to estimate a transverse g factor $\mathrm{\approx 7.5}$, in close agreement with expected transverse g factor of $\mathrm{\approx 8}$ for the g=1.64 transition.
 
To perform echo detected field sweep (EDFS) measurement, we send 1 $\mathrm{\mu s}$ long $\pi$/2 and $\pi$ pulses at a separation of 5 $\mathrm{\mu s}$, and sweep magnetic field to detect spin echo from different transitions (Fig.~1(f)). We observe spin echoes from g=$4.78$, g=$3.28$ in the \textless 150 mT region and g=$1.64$ transitions extending from 160 mT to 275 mT seen as a long tail in the spectra ending around 275 mT.

\section{Spin coherence and relaxation times}

\begin{figure*}[t!]
        \centering
        \includegraphics[width=1\textwidth]{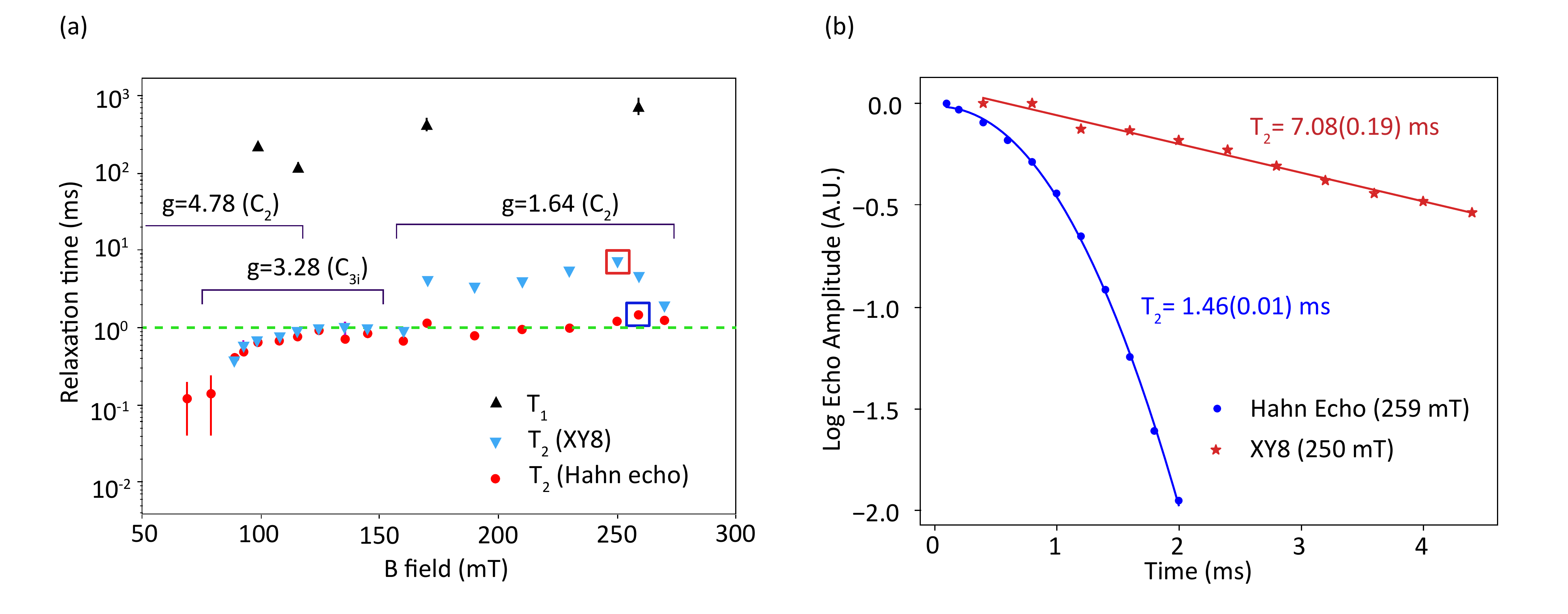}
        % \caption{ Spin lifetime ($\mathrm{T_1}$), Hahn-echo coherence time ($\mathrm{T_2}$) and $\mathrm{XY8}$ dynamic decoupling coherence time measured for  $^{167}\mathrm{Er}^{3+}
% : Y_2 O_3$ transitions}
    \caption{Electron spin coherence and relaxation times on $^{167}\mathrm{Er}^{3+}$: Y\textsubscript{2}O\textsubscript{3} at milliKelvin temperatures. (a) Spin lifetime ($\mathrm{T_1}$) (black dots) was measured using saturation recovery and did not show strong field dependence. Spin coherence time measured with two pulse echo  ($\mathrm{T_2}$,red dots) and with $\mathrm{XY8}$  dynamic decoupling ( $\mathrm{T_2^{\rm XY8}}$, blue dots) showed longer coherence times for lower g transitions. Multiple transition approaching or exceeding 1 millisecond spin coherence time (green dashed line) were measured. Red and blue boxes highlight the longest XY8 and Hahn echo coherence times, respectively. (b) Echo decays for the two transitions with longest Hahn echo and XY8 coherence times, as highlighted in (a).}
\end{figure*}

For the ESR transitions observed in Fig.~1(f), we measure spin relaxation times, Hahn echo coherence times and coherence times under $\mathrm{XY8}$ dynamic decoupling at the base temperature. Fig.~2(a) shows $\mathrm{T_1}$ (black), Hahn echo $\mathrm{T_2}$ (red) and dynamically decoupled $\mathrm{T_2^{\rm XY8}}$ (purple) as a function of B field. We fitted the decay of integrated echo amplitudes with a stretched exponential $\mathrm{exp(-2\tau/T_2)^n}$, where $n$ is the stretch factor. 
Fig.~2(a) shows $\mathrm{T_2}$ coherence times approaching or exceeding 1 ms (green dashed line) for majority of the transitions above 100 mT, with the highest $\mathrm{T_2}$ of 1.46 ms measured around 259 mT from g=1.64 transitions (blue box in Fig.~2(a)). The coherence time increases with fields due to lower sensitivity of high field (lower effective g) transitions to spectral diffusion (SD) and instantaneous diffusion (ID) \cite{pgoldner} noise. At a 5.67 GHz splitting, the sensitivity of ESR transition to magnetic noise $\mathrm{(\partial E /\ \partial B)}$ is simulated to be $\approx \mathrm{g \mu_B}$ (Supplemental information), therefore electron-nuclear spin mixing does not play a significant role in the measured long coherence time \cite{spinechogroundandopter}. The measurable lowest field transition at 70 mT showed a $\mathrm{T_2}$ of 100 $\mathrm{\mu s}$ along with a strong modulation of echo amplitudes due to superhyperfine couplings \cite{superhyperfine} \cite{yizhongandriku}. The stretch factor n increased from 1 for lowest field transition (70 mT) to 2.5 for highest field transition (270 mT), indicating spectral diffusion as a decoherence mechanism for the latter transitions \cite{mimsspectral}.

Decoherence due to spectral diffusion can be suppressed with dynamic decoupling \cite{dynamicdecouplingcolloq}. XY8 sequence in particular offers robust performance for an arbitrary initial state, as well as compensates for pulse errors \cite{xy8paper}. Fig.~2(a) shows a clear increase in coherence times from $\approx$1 ms with Hahn echo to beyond 3 ms with XY8 for high field ($\mathrm{>}$ 150 mT) transitions, with the longest $\mathrm{T_2^{\rm XY8}}$ of 7.08 ms for the 250 mT transition (red box). The echo decays with XY8 sequence were well fitted with a single exponential decay. In this measurement, we sent N= 248 $\pi$ pulses. Further increasing N did not yield longer coherence, indicating an effective filtering of the low frequency noise \cite{siqdotdd}. Therefore, we expect that the measured $\mathrm{T_2^{\rm XY8}}$ is limited by instantaneous diffusion (ID) and possible contribution from higher frequency noise. The coherence time improvement with XY8 was less pronounced for lower field transitions, which suggests an increase in instantaneous diffusion with higher effective g factors, scaling approximately as $\mathrm{g_{eff}^2}$ \cite{idpaper}. This explains the sharp increase in $\mathrm{T_2^{\rm XY8}}$ at around 170 mT. The measured $\mathrm{T_2^{\rm XY8}}$ is in good agreement with the estimated ID limit for g=1.64 transitions of 7 ms (Supplemental information). Spin $\mathrm{T_1}$ for a few transitions was also measured using saturation-recovery sequence and is found to be in the range of 100 ms to 0.7 s (Fig.~2(a) black). 

ESR transitions between 150-250 mT could have mixed contributions from Er and other paramagnetic impurities. Therefore it is imperative to verify those transitions are indeed from Er spins. By illuminating the sample with a laser pulse on resonance with the $\mathrm{C_2}$ site optical transition around 1535.6 nm, we observe a clear optically modulated spin echo signal for the ESR transition around 259 mT, which cannot be attributed to other heating effects (supplemental material), thereby validating its assignment to $^{167}\mathrm{Er}^{3+}$ spins in the $\mathrm{C_2}$ site. 

\section{Decoherence at sub-Kelvin temperatures}

\subsection{Temperature dependence of decoherence and relaxation processes}
 
\begin{figure*}[t!]
        \centering
        \includegraphics[width=1\textwidth]{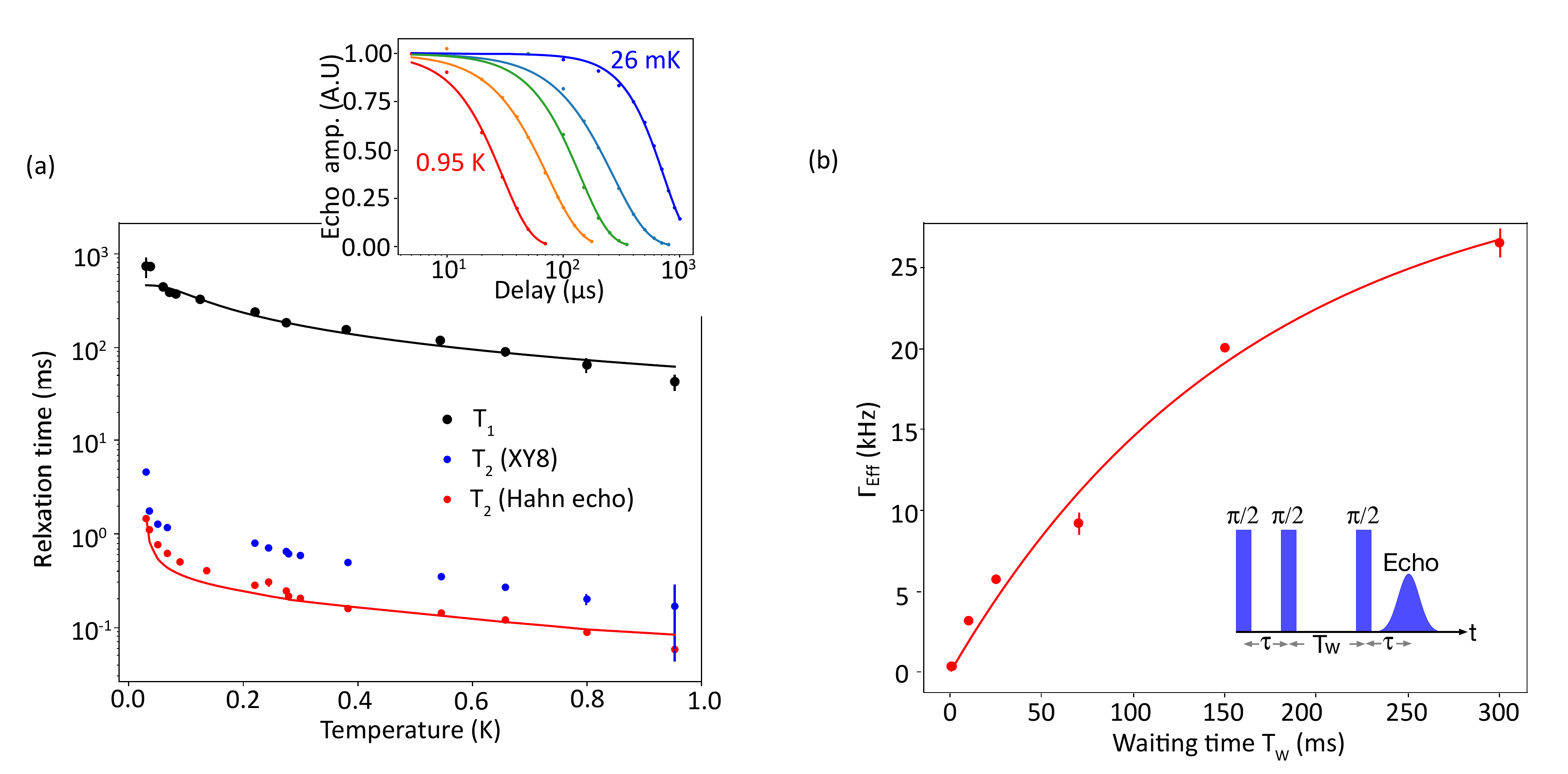}
    \caption{Decoherence mechanisms at sub-Kelvin temperature for $^{167}\mathrm{Er}^{3+}$  g=1.64 (259 mT) transition. (a) Temperature dependence of relaxation and decoherence processes: Spin lifetime ($\mathrm{T_1}$, black dots) was measured using saturation recovery and fitted with $\mathrm{T_1}$ model from Eq.~\ref{T1model} (black line). Coherence time ($\mathrm{T_2}$, red dots) was measured using two pulse echo and fitted with  $\mathrm{T_2}$ model from Eq.~\ref{t2model} (red line). $\mathrm{XY8}$ dynamic decoupling coherence time ($\mathrm{T_2^{\rm XY8}}$) was measured by sending 30 XY8 pulses separated by $\mathrm{2 \tau}$ of 10 $\mathrm{\mu s}$. Inset shows time domain Hahn echo decay for the largest range of temperatures measured. (b) Spectral diffusion measurement with three pulse echo at base temperature of 26 mK. The spin homogeneous linewidth ($\mathrm{\Gamma_{Eff}}$) shows a slow broadening with $\mathrm{\approx}$ 100 ms characteristic timescale with the waiting time between second and third pulse $\mathrm{T_W}$.}
\end{figure*} 

We perform temperature dependent relaxation and coherence time measurements for the 259 mT (g=1.64) transition to understand decoherence mechanisms below 1 K. We use continuous-wave off-resonant laser at 191 THz (1570 nm) to heat the sample directly. The actual sample temperature was estimated from the electron spin echo amplitudes and by correlating it to the recorded stage temperature \cite{milliKelvinESR} (Supplemental material). A sample temperature of 26 mK was determined based on the a spin polarization model when the stage temperature is 7.5 mK. 

The electron spin lifetimes ($\mathrm{T_1}$) measured using saturation recovery (black dots, Fig.~3(a)) show an increase from 42 ms at 0.95 K to 0.73 s at 26 mK. The spin $\mathrm{T_1}$ temperature dependence for rare-earth dopants at temperature below 4K can be modelled by a combination of direct phonon process $\mathrm{R_D}$, the flip-flop $\mathrm{R_{ff}}$ process and a temperature independent component $\mathrm{R_0}$ (black line) \cite{modifcationofphonon}, with $f$ being the transition frequency,

\begin{equation}
    \frac{1}{T_1} = \mathrm{R_0} + \mathrm{R_{ff}  sech^2  \Big( \frac{hf}{2k_BT} \Big)} +  \mathrm{R_{D}  coth  \Big( \frac{hf}{2k_BT} \Big) }
    \label{T1model}
\end{equation}

\noindent The fitted coefficients are $\mathrm{R_D}$= 2.19 Hz, $\mathrm{R_{ff}=}$ 0.87 Hz and  $\mathrm{R_{0}=} 1.67 \times 10^{-8} $ Hz.

The Hahn echo spin coherence times $\mathrm{T_2}$ show strong dependence on temperature (red dots in Fig.~3(a)) from 58 $\mathrm{\mu s}$ at 0.95 K to 1.46 ms at 26 mK. The echo decay was fitted with a stretched exponential and exhibited a stretch factor in the range of 1.3 to 2.1, pointing to spectral diffusion as the dominant decoherence mechanism throughout the temperature range. Improvement in coherence times with XY8 dynamic decoupling ($\mathrm{T_2, (XY8)}$) in the same temperature range is also indicative of a spectral diffusion dominated decoherence (blue dots in Fig.~3(a)).

To model the temperature dependence of $\mathrm{T_2}$ and to obtain spectral diffusion parameters for the environment spin bath, we consider three contributions: spectral diffusion limited coherence time, $\mathrm{T_{2,SD}}$, instantaneous diffusion $\mathrm{T_{2,ID}}$ and spin relaxation limit $\mathrm{2T_1}$,

\begin{equation}
   \mathrm{T_{2}^{-1} = T_{2,SD}^{-1} + T_{2,ID}^{-1} + (2T_{1})^{-1} }  
   \label{t2model}
\end{equation}

Spectral diffusion limited coherence, $\mathrm{T_{2,SD}}$ can be modelled using a Lorentz diffusion model with a stretch factor 2 \cite{idpaper} \cite{pgoldner}:

\begin{equation}
  \mathrm{T_{2,SD}^{-1}}= \frac{\sqrt{\pi \Gamma_{SD} R}}{2}  
  \label{sdt2}
\end{equation}

\noindent where $\mathrm{\Gamma_{SD}}$ represents the spectral diffusion linewidth due to dipolar interaction with environmental spins and R is the flip rate of the environment spins. The temperature dependent $\mathrm{\Gamma_{SD}}$ can be modelled as  \cite{erbiumSDpaper}:

\begin{equation}
\mathrm{\Gamma_{SD} = \Gamma_{Max} sech^2  (g_{env} \mu_B B)/(2k_BT) }
    \label{sdmodel}
\end{equation}

\noindent where $\mathrm{g_{env}}$ is the effective g factor and $\mathrm{\Gamma_{Max}}$ is the maximum dipolar broadening with environment spins causing spectral diffusion. We approximate the total flip rate R(T) for the environment spin bath with the inverse of $\mathrm{T_1}$ from the fit (black curve) in Fig.~3(a). The contribution of decoherence from instantaneous diffusion $\mathrm{T_{2,ID}}$ can be estimated from the measured coherence time after XY8 dynamical decoupling, as it represents the remaining decoherence that cannot be cancelled by the XY8 sequence. 

By fitting the coherence decay, we obtain spin bath parameters $\mathrm{\Gamma_{Max}}$ = $\mathrm{2.80(0.51)}$ MHz and $\mathrm{g_{env}=0.70}$, which are not consistent with $^{167}\mathrm{Er}^{3+}$ spin induced spectral diffusion. While $\mathrm{\Gamma_{Max}}$ of $\mathrm{2.80}$ MHz is significantly larger than the spectral diffusion linewidth of 124 kHz due to erbium spins  (Supplemental material), $\mathrm{g_{env}=0.70}$ is smaller than the lowest g (1.64) of $^{167}\mathrm{Er}^{3+}$. This indicates spin bath is primarily comprised of impurity spins with a low g factor. Indeed we observe a broad ESR signal at 300-900 mT fields (g=1.3-0.4), with peaks at g=0.71 and g=0.81 (Supplemental material). Impurity transitions at fields higher than 900 mT (g=0.4) could not be measured due to the field limit of our magnet. Spins with g=0.4(0.7) are relatively unpolarized even at our lowest measurement temperature, with $\mathrm{87.1(98.1) \%}$ polarization as compared to $\mathrm{99.996 \%}$ polarization for the $^{167}\mathrm{Er}^{3+}$ g=1.64 transition around 259 mT. Therefore, the spectral diffusion measured is attributed to the unpolarized paramagnetic impurities within the host matrix.
 
\subsection{Spectral diffusion at milliKelvin temperatures}

 We study the residual spectral diffusion at the lowest temperature of 26 mK to further validate our model of spectral diffusion. We use three pulse echo (stimulated echo) technique with three $\pi/2$ pulse with the delay between the first and second pulses being $\mathrm{\tau}$ and the delay between second and third pulses being $\mathrm{T_W}$. For each value of $\mathrm{T_W}$ we sweep $\mathrm{\tau}$ to measure the effective linewidth $\mathrm{\Gamma_{Eff} = 1/(\pi T_2)}$ (Fig.~3b):

\begin{equation}
    \mathrm{\Gamma_{Eff} = \Gamma_{0} + \frac{1}{2} \Gamma_{SD} ( 1 - e^{-RT_w}) }
\end{equation}

\noindent where $\mathrm{\Gamma_{SD}}$ is spectral diffusion linewidth, $\mathrm{R}$ is total spin flip rate and $\mathrm{\Gamma_{0}}$ is the linewidth in absence of spectral diffusion \cite{pgoldner} \cite{erbiumSDpaper} .

We obtain $\mathrm{\Gamma_{SD}}$ of $\mathrm{64.5(8)}$ kHz, $\mathrm{R}$ of $5.6(1.5)$  Hz and $\mathrm{\Gamma_{0}}$ of $\mathrm{0.6 (0.8)}$ kHz at 26 mK. 
The fitted $\mathrm{R}$ and $\mathrm{\Gamma_{SD}}$ predict spectral diffusion limited coherence time ($\mathrm{T_{2,SD}}$) of $1.87$ ms from Eq.~\ref{sdt2}, which is in close agreement with the predicted  $\mathrm{T_{2,SD}}$ of $2.2$ ms at 26 mK from the temperature dependent spectral diffusion model.
The total spin flip rate R from three pulse echo $5.6(1.5)$ Hz is higher than the measured $1.4(0.6)$ Hz spin flip rate of $^{167}\mathrm{Er}^{3+}$ g=1.64 transition, further substantiating the presence of paramagnetic impurity spins.

Close agreement between the spectral diffusion measurements from three pulse echo and the temperature dependence model substantiates our understanding of decoherence processes. Decreasing temperature lowers the spectral diffusion linewidth $\mathrm{\Gamma_{SD}}$ by factor of $\mathrm{sech^2  (g_{env} \mu_B B)/(2k_BT)}$, and slows the  total spin flip rate R by freezing the direct phonon ($\mathrm{coth (g_{env} \mu_B B)/(2k_BT)}$) and flip-flop processes ($\mathrm{sech^2  (g_{env} \mu_B B)/(2k_BT)}$). The residual decoherence at 26 mK is dominated by slow spectral diffusion of $\approx$ 5 Hz spin flips and a smaller contribution from instantaneous diffusion (Supplemental material).

To further investigate the spectral diffusion induced by paramagnetic impurities in $\mathrm{Y_2 O_3}$, we perform pulsed ESR on a $\mathrm{Y_2 O_3}$ with no intentional erbium doping and observe spin echo signals from impurity spins in the host matrix (Supplemental material). The $\mathrm{T_2}$ measured on an impurity transition in undoped $\mathrm{Y_2 O_3}$ is comparable to $\mathrm{T_2}$ measured in $^{167}\mathrm{Er}^{3+}$ doped $\mathrm{Y_2 O_3}$ at same field, as expected. This also puts an upper limit on magnetic noise from flip-flops of yttrium nuclear spins ($\mathrm{\delta B_{Y} }$) as $\mathrm{\approx 1/(\pi g \mu_B T_2) =  9 \, nT } $. The $^{167}\mathrm{Er}^{3+}$ transition with 1.46 ms $\mathrm{T_2}$ at 26 mK shows a non-exponential decay with a stretch factor of 2.1, indicating the presence of a frozen core of yttrium nuclei in direct proximity to $\mathrm{Er^{3+}}$ with suppressed flip-flop dynamics \cite{frozencorePhysRevLett.66.695}\cite{Kukharchyk_2018}. While the estimated nuclear spin-bath limited $\mathrm{T_2}$ based on yttrium density in $\mathrm{Y_2 O_3}$ is $\mathrm{\approx}$ 1 ms \cite{doi:10.1073/pnas.2121808119}, we can expect a longer $\mathrm{T_2}$ due to the frozen core effect \cite{Zhong2015} \cite{Kukharchyk_2018}.

We note that spin homogeneous linewidth deviates from linear dependence on temperature as expected in the case of magnetic two-level system (TLS) dominated decoherence in ceramics \cite{yizhongandriku}. We observe a broad feature around g $\mathrm{\approx}$ 2, as well as a sharp g=1.97 signal ($\mathrm{F^+}$ defect center \cite{Kunkel2016}) at high temperature pulsed ESR, witnessing the presence of vacancy defects and magnetic TLS in the host matrix (Supplemental material). However, spins with g $\mathrm{\approx}$ 2 would be frozen with 99.9997 \% polarization at 26 mK and 259 mT, and are not expected to contribute significantly to spectral diffusion \cite{multimodeeryso}. % does magnetic TLS polarize following same eqn as spin?

\begin{figure*}[t!]
    \centering
     \includegraphics[width=1\textwidth]{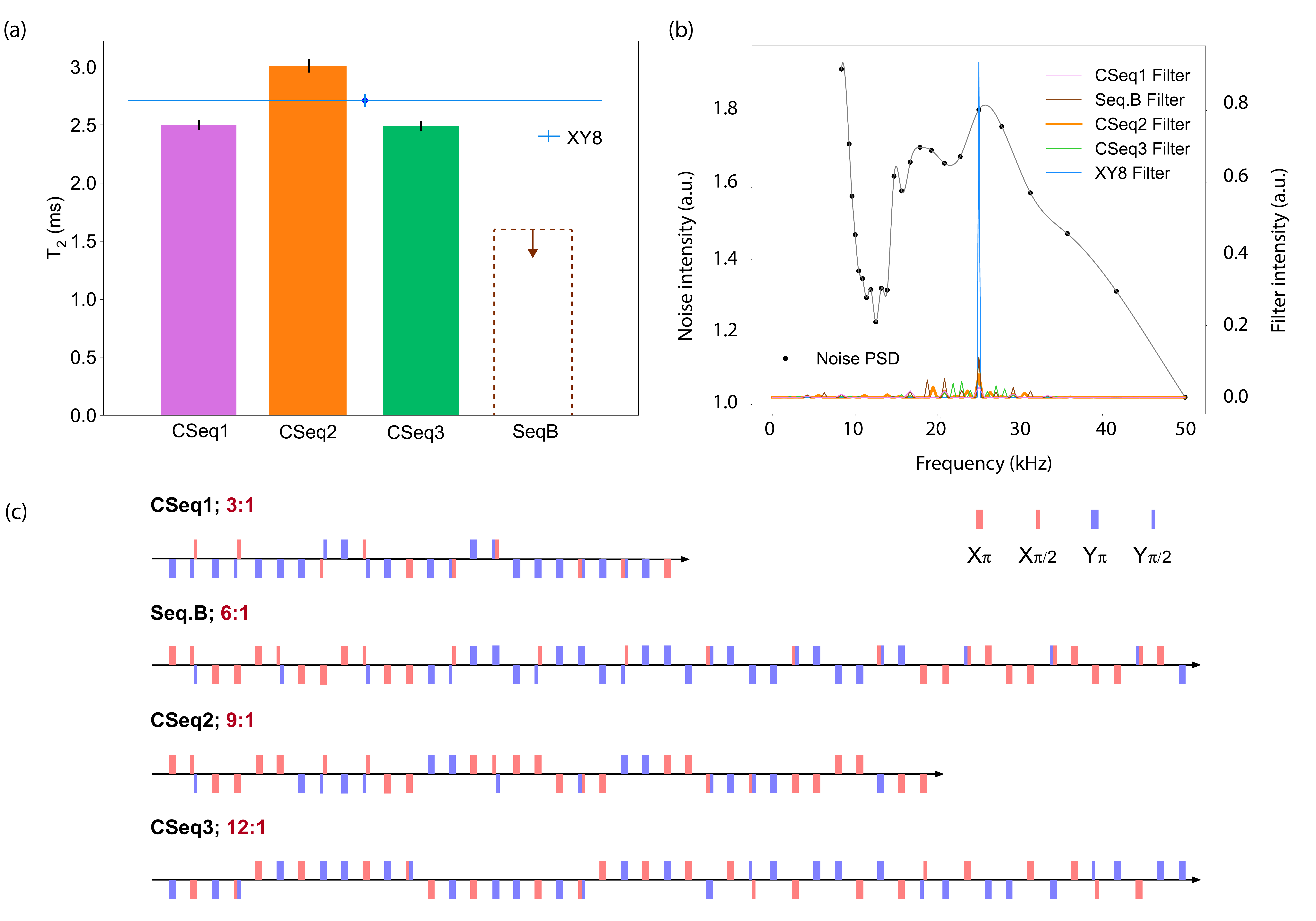}
    \caption{Coherence enhancement with customized dynamic decoupling. (a) Comparison of coherence times under different dynamic decoupling sequences at 25 kHz center filter frequency. CSeq2 outperforms all customized dynamical decoupling sequences and XY8. $\mathrm{T_2}$ for Seq.B is not able to measure due to fast decaying spin echo below the noise floor. (b) Noise spectrum computed using CPMG $\mathrm{T_2}$ measurement and comparison of the filter functions of all sequences. Filters are normalized in the whole spectrum but only 0-50kHz range is shown given the measured noise spectrum range. (c) Customized dynamic decoupling pulse sequences in increasing order of the target ratio of disorder to interaction, along with Seq.B.} \label{fig-custom}
\end{figure*}

\section{Coherence time enhancement beyond spectral diffusion limit}

The spin coherence time measurements for g=1.64 transitions in Fig.~2(a) show instantaneous diffusion by dipolar interactions in the Er spin ensembles. This decoherence cannot be cancelled by standard dynamic decoupling sequences like XY8. A new sequence is desired so that it can mitigate both spectral diffusion and dipolar interactions with robustness to pulse errors at the same time \cite{hamitonianenging}. A previously proposed sequence for disorder-dominated sequence (Seq.B) has shown improvement over XY8 dynamic decoupling for electron spins in NV centers \cite{quantummetrology} and Er:YSO \cite{merkel2020dynamical}. These systems, however, had strong dipolar interactions with XY8 dynamically decoupled $\mathrm{T_2}$ of $\mathrm{\approx 1 \mu s}$ as compared to $\mathrm{T_2}$ of $\mathrm{\approx ms} $ in our system. For systems with weaker dipolar interactions, customized sequences are thus required. In this section, we describe the results of three customized sequences --- CSeq1 for 3:1 (ratio between spectral diffusion and interaction decoupling), CSeq2 for 9:1 and CSeq3 for 12:1, together with existing XY8 (inf.:1) and Seq.B (6:1) on the 259 mT (g=1.64) transition. We note that the ratio assigned to a sequence is how fast it decouples from slow-varying spectral diffusion noise versus dipolar interaction induced broadening, and it is an intrinsic property of the sequence. The actual ratio of spectral diffusion to dipolar interaction noise experienced by the spin system is difficult to predict when there is noise fluctuating faster than the sequence time scale. Therefore, it is important to design sequences with varying ratios to better match the actual noise characteristics of the system.

The starting point for the design of customized sequences is that a given sequence features different cycle lengths to mitigate either spectral diffusion or dipolar interactions by moving the system into a 'Toggling Frame'\cite{hamitonianenging}, where we can obtain a pictorial visualization of how the original spin Pauli operators evolve under the Floquet engineered version~\cite{U.Haeberlan;J.S.Waugh2015}. Figure ~4(a) presents coherence time measurements for different sequences centered at a 25kHz filter frequency. We observe that one customized sequence --- CSeq2 outperforms all other sequences and even the XY8 sequence. All customized sequences yielded longer coherence than Seq.B, with CSeq2 reached a nearly factor of 2 improvement over Seq.B. We also note that when measuring Seq.B, spin echo amplitudes quickly diminished below the noise floor after a few data points. Therefore, only an upper limit of 1.6 ms was placed on Seq.B $\mathrm{T_2}$.

To understand the performance variations among difference sequences, we calculate their effective filter functions and compare with that of XY8 in Fig.~4(b). We also plot the noise power spectrum density (PSD) experienced by the system obtained by CPMG (Carr-Purcell-Meiboom-Gill) sequences with varying pulse separations (detailed information in the supplemental material). While the dominant source of spectral diffusion is low frequency noise with $\mathrm{\approx}$ Hz flip rate as discussed in section IV, the noise PSD also exhibits a local maxima around 25 kHz. Compared with the XY8 sequence, the filter functions of the customized sequences generate sidebands which will sample less noises in total. We see from Fig.~4(b) that some customized sequences have wider sidebands than CSeq2 (i.e. CSeq1 and Seq.B) but do not outperform CSeq2 in terms of the coherence time, indicating that the amount of dipolar interaction suppressed vary among different sequences and the coherence enhancement is not solely due to the spread of the filter function. Nonetheless, due to the fact that the electronic spin g-tensor is highly anisotropic in Er$^{3+}$ ions, customized sequences cannot fully cancel the dipolar interaction term in the spin Hamiltonian~\cite{merkel2020dynamical}. In this case, we can only tune the ratio between transverse ($J_{S}$) and longitudinal ($J_{I}$) coupling in the spin system by rearranging the pulse separations in the sequence, which give rise to relaxation and dissipation, respectively~\cite{quantummetrology}. A longer coherence beyond current best result remains possible if such ratio can be fine tuned. However, a full optimization was not performed in this work due to limitations of our experimental setup, such as the smallest accessible pulse width ($\mathrm{1 \,\, \mu s}$) and the minimum pulse separation ($\mathrm{10 \,\, \mu s}$).

\section{Discussion}

Our results show milliseconds erbium spin coherence is achievable despite the presence of nuclear spins (e.g. yttrium) and paramagnetic impurities in the host matrix. To generalize, we obtain a long spin coherence time by addressing low g-factor transitions and by freezing paramagnetic impurity spins at moderate fields and milliKelvin temperatures. The remaining decoherence is attributed to dipolar interactions in the weakly interacting Er spin ensembles, which can be further mitigated by custom designed dynamical decoupling sequence to simultaneously suppress spectral diffusion and dipolar interactions. This set of methods are generally applicable to erbium or other Kramers ions in wide-ranging hosts with low site symmetries and anisotropic g tensors. According to \cite{Ammerlaan2001}, a plethora of crystalline hosts for Er in trigonal, tetragonal, monoclinic and orthohombic crystal field symmetries exhibit a lowest component of g $\mathrm{<}$ 2. A subset of them are also low noise hosts with nuclear spin magnetic moments comparable or less than yttrium, such as $\mathrm{CaWO_4}$ (tetragonal),  $\mathrm{SrWO_4}$ (tetragonal), ZnS (tetragonal), MgO (tetragonal), Si (monoclinic), $\mathrm{Y_2 Si O_5}$ (monoclinic) \cite{Ammerlaan2001} \cite{doi:10.1073/pnas.2121808119}. In these hosts, millisecond erbium spin coherence times are expected following our strategy.

Our work along with a similar report in $^{167}\mathrm{Er}^{3+}$:$\mathrm{Ca WO_4}$ \cite{https://doi.org/10.48550/arxiv.2203.15012} shows low g paramagnetic impurities as the dominant source of spectral diffusion in moderately doped ($\mathcal{O}\mathrm{(10)}$ ppm) hosts at milliKelvin temperatures.  Reaching nuclear spin limited coherence time would require hosts with higher chemical purity and a very low concentration of paramagnetic dopants and defects on the $\mathrm{\approx}$ ppb level \cite{doi:10.1126/sciadv.abj9786}. Nevertheless, for applications based on moderately doped rare-earth ensembles (($\mathcal{O}\mathrm{(10)}$ ppm)), such as optical quantum memories and quantum transduction, either dipolar interactions or paramagnetic impurities are likely the coherence limiting factors, rather than the nuclear spin limits. In such cases, hosts with weak nuclear spins can provide coherence time on par with nuclear spin free hosts.

The reduced magnetic noise environment also indicates $^{167}\mathrm{Er}^{3+}$ nuclear spin $\mathrm{T_2}$ in excess of a second is achievable in $\mathrm{Y_2 O_3}$ at similar conditions. With the known hyperfine interactions \cite{y2o3esr2022}, the $^{167}\mathrm{Er}^{3+}$ nuclear spins can be initialized into one hyperfine level by sequential hyperpolarization via the Er electron spin, or via optical pumping \cite{Rancic2018}. This promising combination of long-lived electron-nuclear spins and hyperfine couplings them will enable multi-qubit quantum nodes to be capable of entanglement generation, storage and distillation.

\section{Conclusion}
We measured $^{167}\mathrm{Er}^{3+}$ electron spin coherence time exceeding 1 ms in $\mathrm{Y_2 O_3}$ matrix without addressing ZEFOZ transitions. By freezing a spin bath composed of low g paramagnetic impurities with a moderate ($\mathrm{<}$ 1T) field at millikelvin temperatures, a $\mathrm{T_{2}}$ of 1.46 ms for lowest g (1.64) transition was achieved. With XY8 dynamical decoupling, we measured up to 7.1 ms coherence time, among the longest-lived erbium electron spins in nuclear spin matrix with similar doping concentrations. The remaining decoherence was attributed to Er-Er dipolar interactions, which can be further mitigated by custom designed sequence when both spectral diffusion and dipolar interactions are present. We successfully demonstrated the effect of simultaneous suppression of diffusion and interaction with our ratio-targeted dynamical decoupling sequence, which can outperform traditional (XY8) sequence under specific noise configurations. Further optimization is expected with improved experimental setups.

Our work demonstrates that milliseconds erbium electron spin coherence time can be reliably achieved, following our methods, in a wide variety of hosts including those containing nuclear spins. For applications such as quantum memory, hosts with nuclear spin can be advantageous by offering nuclear spin registers for long-term store of quantum information \cite{Ruskuc2022}.  A long-lived Er electron spins along with telecom C-band optical transition and available long-storage nuclear spins makes $^{167}\mathrm{Er}^{3+}$ a significant contender for quantum technologies such as telecom spin-photon interfaces and quantum memories.

\begin{acknowledgments}
\noindent We thank Tijana Rajh for useful discussions, and Ruoming Peng for valuable suggestions to the manuscript. This work was supported by National Science Foundation faculty early career development program (CAREER) grant no. 1944715, the U.S. Department of Defense, Army Research Office grant no. W911NF2010296, the U.S. Department of Energy, Q-NEXT quantum science and engineering research center, and the NSF QLCI for Hybrid Quantum Architectures and Networks (NSF Grant No. 2016136).

\end{acknowledgments}

% \nocite{*}

% \bibliography{apssamp}% Produces the bibliography via BibTeX.

\newpage

 \bibliography{apssamp}

\end{document}